# The non-equilibrium temperature beyond local equilibrium assumption

Zheng-Chuan Wang

The University of Chinese Academy of Sciences, P. O. Box 4588, Beijing 100049, China.

wangzc@ucas.a.cn

Abstract

In this manuscript, we propose a non-equilibrium temperature by a temperature dependent Vlasov equation for the charge particles transport through a environmental reservoir. A new damping force and a inverse damping relaxation time are derived based on the Vlasov equation, which have obvious influence on the external force and the relaxation time of transport particles. The non-equilibrium temperature for the transport particle is defined by it's distribution function out of equilibrium, which is different from the equilibrium temperature of reservoir. There exists heat transfer between the transport particles and the reservoir, because the whole transport particles are in non-equilibrium state. Finally, we illustrate them by an example of one-dimensional charge particles transport under an external electric field, the non-equilibrium temperature and damping force defined by us are shown numerically.

PACS: 05.20.Dd; 05.60.cd; 05.70.ln

# I. Introduction

As we know, the usual temperature is defined for the equilibrium system in thermodynamics theory, the temperature in the non-equilibrium system is always introduced by the local equilibrium assumption[1]. However, this local equilibrium assumption is only suitable for describing the system near equilibrium, and not applicable to the system far away from equilibrium, we can not adopt the local equilibrium assumption for the latter at all. So how to define the temperature in the system far away from equilibrium instead of using local equilibrium assumption is still an open question[2]. On the other hand, Hartmann et al. further pointed out that the temperature will be ill-defined for the nano-scale system[3], they gave a minimum length for the size of this system in which the number of atoms is not macroscopic number demanded by thermodynamics. The energy distribution will deviate substantially from the canonical one below this minimum length, then the temperature is not easy to define, because there is no one-to-one mapping between temperature and the expectation value of physical observables by which the temperature is measured[4].

In this manuscript, we try to define the non-equilibrium temperature for the particles by the distribution function far away from equilibrium which satisfies the Boltzmann equation. As a powerful tool to investigate the classical transport problems, Boltzmann equation was first established in the nineteenth century by Boltzmann in a dilute gas[5]. In the latter, it was extended to study the transport procedures concerning with conduction electrons in metals and semiconductors etc., even to the quasi-particle or elementary excitation in a strongly interacting system, while the latter can be described by the Boltzmann equation because the interaction between quasi-particle is weakly[6]. In order to consider the quantum transport in small systems with obvious quantum effects, Wigner proposed a quantum Boltzmann equation satisfied by the Wigner distribution function which is transformed from the non-equilibrium Green function by a quantum Wigner transformation[7,8]. To study the spin-polarized transport in spintronics, the spin freedom was further included into quantum Boltzmann equation, then the so-called spinor Boltzmann equation[9,10] were obtained, which can also be derived by the non-equilibrium Green function theory similar to Mahan's method[11,12].

If we expand the non-equilibrium distribution function around the local equilibrium distribution, which is called the local equilibrium assumption[1], one can introduce the temperature of the local equilibrium distribution into Boltzmann equation in order to further explore the thermal or thermo-electric transport in the system. Based on this assumption, one can also expand the Wigner distribution function in quantum Boltzmann equation or the spinor distribution function in spinor Boltzmann equation around the local equilibrium distribution, and investigate the thermo-electric transport procedure by use of them. For the spinor Boltzmann equation, we can investigate not only the thermo-electric transport but also the thermal spin transfer torque by the local equilibrium distribution[13], even the thermal spin-orbit torque for the spin switching driven by electric current[14].

However, the local equilibrium assumption is not suitable for the system far away from equilibrium, so how to define the temperature in non-equilibrium system instead of using local equilibrium assumption is interesting. In this manuscript, we introduce the temperature into Boltzmann equation by the reservoir interacting with the system instead of the local equilibrium assumption for the transport particles, in which the distribution function for the particles of reservoir is chosen as the equilibrium distribution function, while the transport particles which collide with the particles of reservoir are still far away from equilibrium, we don't adopt the local equilibrium assumption for them at all. By this way, we obtain a temperature dependent Boltzmann equation, which contains two new temperature dependent terms: one is a damping force, the other is a inverse damping relaxation time. We will show them in the next section.

## II. Theoretical formalism

Consider the transport of a kind of charge particles with mass $m$ through environment whose particles have no charge but with mass $m_1$, the environment can be regarded as a reservoir. There exist collisions between the charge transport particles and the neutral environmental particles, the collisions are short distant and the collision integral in Boltzmann equation is usually written as：

$$C(f,f_T) = \iint [f(\vec{r},\vec{v}',t)f_T(\vec{r},\vec{v}'_{1T},t) - f(\vec{r},\vec{v},t)f_T(\vec{r},\vec{v}_{1T},t)]g_T\sigma_T(\chi_T,g_T)d\vec{v}_{1T}d\vec{k}'_T \qquad (1)$$

where $f(\vec{r},\vec{v},t)$ and $f_T(\vec{r},\vec{v}_{1T},t)$ are the distribution functions of the charge transport particle and the environmental particle, respectively. $\sigma_T(\chi_T,g_T)$ is the differential cross section of collision, where $\vec{g}_T = \vec{v} - \vec{v}_{1T} = g_T\vec{k}_T$ is the relative velocity of two particles with $\vec{v}$ and $\vec{v}_{1T}$ before collision, $g_T = |\vec{v} - \vec{v}_{1T}|$ is it's magnitude, while $\vec{k}_T$ is it's unit vector. Similarly, $\vec{g}'_T = \vec{v}' - \vec{v}'_{1T}$ is the relative velocity of them after collision, $\chi_T$ is the angle between relative velocities $\vec{g}_T$ and $\vec{g}'_T$, or the unit vectors $\vec{k}_T$ and $\vec{k}'_T$. Since the coulomb interaction between charge transport particles is long distant, under the mean field approximation a charge transport particle can be regarded as moving in a mean electric field provided by other charge transport particles, where the mean electric field obeys the PÖisson equation. Along with the Boltzmann equation, the transport particles satisfy the following so-called Vlasov equation[15]

$$\frac{\partial f(\vec{r},\vec{v},t)}{\partial t} + \vec{v} \cdot \vec{\nabla} f(\vec{r},\vec{v},t) + \vec{F}(r) \cdot \vec{\nabla}_v f(\vec{r},\vec{v},t) + \frac{e}{m} \vec{\nabla} \varphi \cdot \vec{\nabla}_v f(\vec{r},\vec{v},t) = C(f, f_T)$$

$$\nabla^2 \varphi = \frac{e}{\varepsilon_0} \int f_1 \tag{2}$$

where $\vec{F}(r) = -\frac{e}{m}\vec{E}$ is the external electric field force applied to the charge transport particles, $\varphi$ is the self-consistent potential of mean electric field provided by other charge particles, $\varepsilon_0$ is the dielectric function of vacuum, $f_1$ is the deviation of distribution function from the equilibrium distribution $f_0$, $f_1 = f - f_0$. We ever extended the Vlasov equation to the case of spinor Boltzmann equation[16].

Although the transport charge particles are out of equilibrium, the environmental particles can still be assumed to remain in the equilibrium, because the relaxation time for the environmental particles is very quick compared with the transport particles, so the environment can be regarded as a heat reservoir with a temperature $T$, it's distribution function can be simply approximated as the equilibrium distribution function, then it's temperature can be naturally introduced into the Boltzmann equation by the collision term with transport particle.

Since the reservoir is very large, the particles in it will exchange the energy and momentum by collisions rapidly, we can assume the particles in the reservoir to keep in the local equilibrium with a temperature $T(x)$, then it's distribution can be expressed as

$$f_T(\vec{r},\vec{v}_{1T},t) = n(\frac{m_1}{2\pi k_B T(x)})^{3/2} e^{-\frac{m_1}{2\pi k_B T(x)}[(\vec{v}_{1T}-\vec{v}_{1T0})^2]} \qquad (3)$$

where $n$ is the density of reservoir particles, $\vec{v}_{1T0}$ is the velocity of mass center, $k_B$ is the Boltzmann constant. Considering the conservation of momentum during the collision, $g_T d\vec{k}_T^{'}$ in the collision term $C(f,f_T)$ can be expressed as $g_T d\vec{k}_T^{'} = (1+\frac{m}{m_1})d\vec{v}^{'}$, then the collision term $C(f,f_T)$ can be written as

$$C(f,f_T) = \int d\vec{v}^{'} \int d\vec{v}_{1T} [f(\vec{r},\vec{v}^{'},t)f_T(\vec{r},\vec{v}^{'}_{1T},t) - f(\vec{r},\vec{v},t)f_T(\vec{r},\vec{v}_{1T},t)]\sigma_T(\chi_T,g_T)(1+\frac{m}{m_1}) \qquad (4)$$

Since the relaxation procedure of particles in the reservoir are very fast, we can adopt Markov approximation on the distribution function of particles in reservoir, which means the particles of reservoir will lose their memory, their velocities $\vec{v}^{'}_{1T}$ and $\vec{v}_{1T}$ have no relation with the velocities $\vec{v}^{'}$ and $\vec{v}$ of the transport particles despite of the collision, which is similar to the molecular chaos assumption adopted during the derivation of Boltzmann equation from BBGKY hierarchy[1,2], then the above collision integral can be approximated as

$$C(f,f_T) = \int d\vec{v}^{'} \{f(\vec{r},\vec{v}^{'},t) \int d\vec{v}_{1T} \left[f_T(\vec{r},\vec{v}^{'}_{1T},t)\sigma_T(\chi_T,g_T)\left(1+\frac{m}{m_1}\right)\right] -$$

$$f(\vec{r},\vec{v},t) \int d\vec{v}_{1T} \left[f_T(\vec{r},\vec{v}_{1T},t)\sigma_T(\chi_T,g_T)\left(1+\frac{m}{m_1}\right)\right]\} \qquad (5)$$

Similar to the Moyal expansion in the derivation of Fokker-Planck

equation from master equation[1,2], we can make the corresponding expansion on the collision integral. If $\Delta \vec{v} = \vec{v}' - \vec{v}$ is very small, $f(\vec{r},\vec{v}',t)$ and $f_T(\vec{r},\vec{v}'_{1T},t)$ in the first term can be expanded by a Taylor series around $f(\vec{r},\vec{v},t)$ and $f_T(\vec{r},\vec{v}_{1T},t)$, it's zero order term in the expansion can be canceled with the second term in Eq.(5), then we have

$$C(f, f_T) = \sum_{n=1}^{\infty} \frac{(-1)^n}{n!} \frac{\partial^n}{\partial \vec{v}^n} [\alpha_n(\vec{v}) f(\vec{r},\vec{v},t)] \tag{6}$$

where

$$\alpha_n(\vec{v}) = \int (\vec{v}' - \vec{v})^n [\int d\vec{v}_{1T} [f_T(\vec{r},\vec{v}_{1T},t) \sigma_T(\chi_T, g_T) \left(1 + \frac{m}{m_1}\right)] d\vec{v}' \tag{7}$$

Since the higher order terms in the Taylor series expansion are small and can be neglected, we only keep the first and second order terms in the equation, then we arrive at

$$\frac{\partial f(\vec{r},\vec{v},t)}{\partial t} + \vec{v} \cdot \vec{\nabla} f(\vec{r},\vec{v},t) + (\vec{F}(r) + \frac{e}{m} \vec{\nabla} \varphi) \cdot \vec{\nabla}_v f(\vec{r},\vec{v},t) - \vec{\nabla}_v [\alpha_1(\vec{v}) f(\vec{r},\vec{v},t)] +$$
$$\frac{1}{2} \vec{\nabla}_v^2 [\alpha_2(\vec{v}) f(\vec{r},\vec{v},t)] = 0 \tag{8}$$

where

$$\vec{\alpha}_1(\vec{v}) = \int (\vec{v}' - \vec{v}) [\int d\vec{v}_{1T} [f_T(\vec{r},\vec{v}_{1T},t) \sigma_T(\chi_T, g_T) \left(1 + \frac{m}{m_1}\right)] d\vec{v}' \tag{9}$$

and

$$\alpha_2(\vec{v}) = \int (\vec{v}' - \vec{v})^2 [\int d\vec{v}_{1T} [f_T(\vec{r},\vec{v}_{1T},t) \sigma_T(\chi_T, g_T) \left(1 + \frac{m}{m_1}\right)] d\vec{v}' \tag{10}$$

As pointed out in the above, the environmental particles can be regarded as remain in the equilibrium with a equilibrium distribution function at certain temperature, then the coefficients $\vec{\alpha}_1(\vec{v})$ and $\alpha_2(\vec{v})$ will be temperature dependent. If we only keep the first order term in

Eq.(8), the temperature dependent Boltzmann equation (8) can be further simplified as

$$\frac{\partial f(\vec{r},\vec{v},t)}{\partial t} + \vec{v}\cdot\vec{\nabla}f(\vec{r},\vec{v},t) + \left(\vec{F}(r) + \frac{e}{m}\vec{\nabla}\varphi - \vec{\alpha}_1(\vec{v})\right)\cdot\vec{\nabla}_v f(\vec{r},\vec{v},t) = \left(\vec{\nabla}_v\cdot\vec{\alpha}_1(\vec{v})\right)f(\vec{r},\vec{v},t) \quad (11)$$

Usually, the collision integral in the Boltzmann equation can be simplified by the relaxation time approximation $-\frac{f-<f>}{\tau}$, where $\tau$ is the relaxation time, so the term in the right hand of Eq.(11) corresponds to the relaxation term except a constant term $\frac{<f>}{\tau}$, so the velocity derivative of $\vec{\alpha}_1(\vec{v})$ has influence on the relaxation time of transport particles, which modifies the constant relaxation time $\tau$ as velocity dependent $\tau(\vec{v})$ as shown in Ref.[17], we call it the inverse damping relaxation time,

$$\tau(\vec{v}) = \frac{1}{\vec{\nabla}_v\cdot\vec{\alpha}_1(\vec{v})}. \quad (12)$$

While the temperature dependent term $\vec{\alpha}_1(\vec{v})$ brings a correction on the external force, too, we term it as the temperature dependent damping force. Both of the damping force and inverse damping relaxation time come from the collision between the transport particles and environmental particles. So we finally obtain the temperature dependent Vlasov equation.

It should be emphasized that the temperature in Eq.(11) is still an equilibrium temperature, it is just the equilibrium temperature of the environmental reservoir. But we haven't adopted the local equilibrium assumption for the transport particle at all, because it is far away from the equilibrium due to the big external force. In fact, we can define a non-equilibrium temperature for the transport particles. To explore this issue, we try to use the non-equilibrium distribution function of transport particles, and define

$$\frac{3}{2}k_B T_{neq}(x,t) = <\frac{1}{2}m(\vec{v}-\vec{v}_0)^2> = \int \frac{1}{2}m(\vec{v}-\vec{v}_0)^2 f(\vec{r},\vec{v},t)d\vec{v} \tag{13}$$

Obviously, it is different from the equilibrium temperature of reservoir, it is position and time dependent, we haven't adopted the local equilibrium assumption in the distribution function of transport particles at all. To see the difference between the non-equilibrium temperature $T_{neq}(x,t)$ and equilibrium temperature clearly, we study charge particles transport under a small external electric field. At steady state, it's distribution function can be expanded by the small external electric field, it is

$$f = f_0 + f_1 + f_2 + \dots \tag{14}$$

where $f_0$ is the equilibrium distribution function, and $f_1 = \frac{q\tau}{\hbar}\vec{E}\cdot\vec{\nabla}_k f_0$, $f_2 = \frac{q\tau}{\hbar}\vec{E}\cdot\vec{\nabla}_k f_1$.... For simplicity, we only keep the first order term and substitute it into Eq.(13), we have

$$\frac{3}{2}k_B T_{neq}(x,t) = \frac{3}{2}k_B T_{eq} + \int \frac{1}{2}m(\vec{v}-\vec{v}_0)^2 \frac{q\tau}{\hbar}\vec{E}\cdot\vec{\nabla}_k f_0 d\vec{v}, \tag{15}$$

where $\frac{3}{2}k_B T_{eq} = \int \frac{1}{2}m(\vec{v}-\vec{v}_0)^2 f_0 d\vec{v}$. It is shown that the non-equilibrium temperature is determined not only by the equilibrium temperature $T_{eq}$, but also by the external electric field $\vec{E}$, the latter will increases the non-equilibrium temperature $T_{neq}$. Certainly, Eq.(15) is only applicable for the small external electric field, but Eq.(11) is general whatever for big or small external electric field. Since the temperature of charge transport particles and the temperature of reservoir are different, there exist heat transfer between them, the whole system remain in non-equilibrium.

**III. Numerical Results**

As an example, for simplicity we consider the transport of charge particles, i.e., the ions under a big external electric field, there exist another kind of particles without charge in the environment, which can be regarded as reservoir. Suppose the environmental particles with mass $m_1$ collide with the charge particles by the short distant interaction, and we further assume the particles in reservoir remain in the local equilibrium with a distribution like Eq.(3), in which the particle density $n$ can be expressed as

$$n(x) = n_0 e^{\frac{m_1}{2k_B T(x)} v_{0T}^2 - \frac{m_1}{k_B T(x)} \phi(x)} \tag{16}$$

where $n_0$ is a constant, $T(x)$ is the position dependent temperature of reservoir, $\vec{v}_{0T}$ is the velocity of mass center, we simply choose it to be zero, $\phi(x)$ is the potential of gravity field felt by the particles of reservoir. Since we have applied a big external electric field to the charged transport particles, we can simply neglect the smaller mean electric field $\vec{\nabla}\varphi$ in Eq.(11) provided by other charged particles. In our calculation, we chose $T(x) = T_0$ as a constant, which means the reservoir remain in the equilibrium. The electric field is adopted as

$E = 0.1 mV/nm$, which is big enough to push the system far away from equilibrium.

In Fig.1, we plot the temperature dependent damping force as a function of position $x$ at different equilibrium temperatures $T_0$, it can be seen that the damping force increases gradually with position, it's magnitude is comparable to the external electric field force, so it can not be neglected. The damping force has a negative value, which will weaken the effect of external electric field force. The damping force also depends on the temperature of reservoir, in Fig.1 the dashed line corresponds to the case at equilibrium temperature $T_0 = 180K$, while the solid line corresponds to $T_0 = 410K$, it is shown that the absolute magnitude of damping force increases with the temperature of reservoir, which is conceded to it's definition in Eq.(9). In Fig.2, we also draw the inverse damping relaxation time as a function of position at different temperatures $T_0$, it indicates that the inverse damping relaxation time increases nearly linearly with position. When the temperature of reservoir increases from $T_0 = 180K$ to $T_0 = 410K$, the absolute magnitude of inverse damping relaxation time becomes bigger.

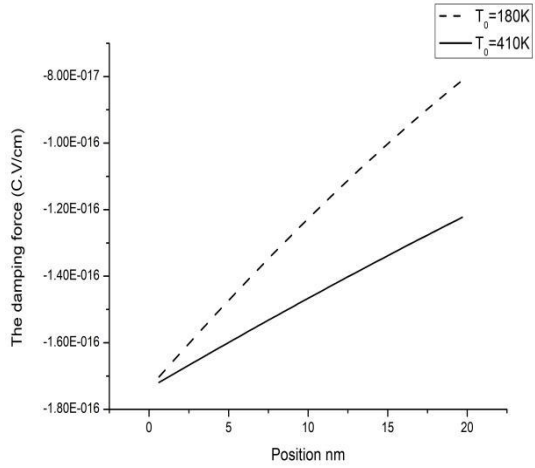
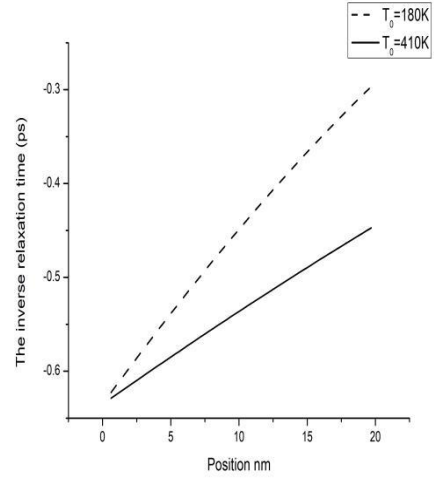

Fig.1 The damping force vs position, where $E = 0.1mV/nm$.

Fig.2 The inverse damping relaxation time vs position, where $E = 0.1mV/nm$.

To demonstrate the difference of non-equilibrium temperature $T_{neq}(x,t)$ for transport particles and the equilibrium temperature $T_0$ for environmental particles, we show the non-equilibrium temperature in fig.3 in the case of $T_0 = 180K$, it is position dependent and decreases with position. Since the external electric field will increases the non-equilibrium temperature according to Eq.(15), the latter has a value bigger than the temperature of reservoir $T_0 = 180K$. There exist heat transfer from the transport particles to reservoir due to the scattering between them. The difference of non-equilibrium temperature and equilibrium temperature of reservoir is obvious, which means the transport particles are not in the equilibrium state due to the external electric force. In fig.4, we also plot the charge current density as a function of position, it has a similar shape with non-equilibrium temperature, because they are all calculated by the non-equilibrium distribution function which has the similar shape. The charge current density decreases with position due to the scattering of electrons with particles of reservoir.

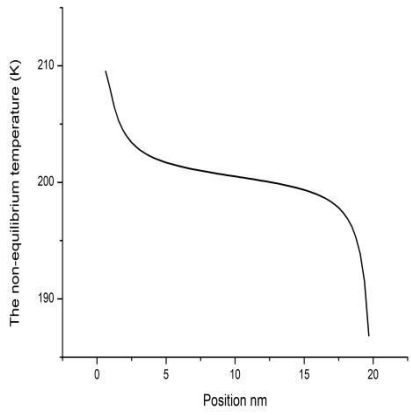 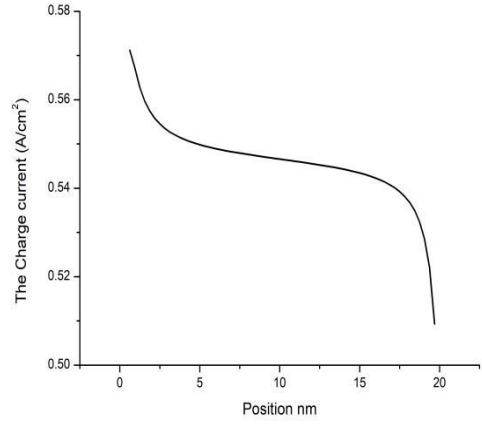

Fig.3 The non-equilibrium temperature vs position, where $E = 0.1 mV/nm$, $T_0 = 180K$

Fig.4 The charge current vs position, where $E = 0.1 mV/nm$, $T_0 = 180K$.

## IV. Summary and discussions

In summary, we proposed a non-equilibrium temperature by a temperature dependent Vlasov equation for the system with two kinds of particles: the charge transport particles and the environmental particles, while the latter are assumed remain in the equilibrium. The new damping force and inverse damping relaxation time are obtained based on the Taylor series expansion on the collision integral. This temperature in Boltzmann equation comes from the temperature of reservoir which is assumed to be in local equilibrium. We define the non-equilibrium temperature by the distribution function of transport particle which is far away from equilibrium driven by the big external force, it is different from the equilibrium temperature of reservoir, there exist heat transfer between transport particles and reservoir. The temperature dependent damping force and inverse damping relaxation time are also evaluated numerically, we show them in Fig.1 -2, while the non-equilibrium temperature and charge current density as a function of position are demonstrated in Fig.3 and Fig.4, respectively.

It should be noted that the particles of reservoir in our manuscript have their own mass, in fact there are many kinds of reservoirs, the particles of some reservoirs own their mass, while others maybe have no mass, i.e., the radiative photons. If the reservoir are composed by radiative photons, the collision integral will not be written as Eq.(1) in our manuscript, it should take the form of Compton scattering, we have discussed them in a conference paper[18].

Meanwhile, for simplicity, we only consider the collision of charged transport particle with neutral environmental particle. In fact, in a conductor the collision usually happen between different charged particles, i.e., the conduction electrons and the charged background particles (the ionic cores in the lattice), then the collision term can not be described by the integral in Eq.(1) which is only suitable for the description of collision with short distant interaction, it is complicated. In this case the damping force and the inverse damping relaxation time will be different correspondingly, we leave it for future exploration.

**Acknowledgments**

This study are supported by the National Key R&D Program of China (Grant No. 2022YFA1402703), the Key Research Program of the Chinese Academy of Sciences (Grant No. XDPB08-3).

**Data Availability Statement**

Data sets generated during the current study are available from the corresponding author on reasonable request.

**Additional information**

Competing interest statement: The authors declare that they have no competing interests.